\documentclass[twocolumn,rotate,aps,prl,showpacs,amsmath,nobalancelastpage]{revtex4}
\usepackage{amssymb}
\usepackage{graphicx}

\begin{document}

\title{Polymer heat transport enhancement in thermal convection:
the case of Rayleigh-Taylor turbulence}

\author{G. Boffetta$^{1}$, A. Mazzino$^{2}$, S. Musacchio$^{3}$  
and L. Vozella$^{2}$}
\affiliation{$^1$Dipartimento di Fisica Generale and INFN, Universit\`a di Torino,
via P.Giuria 1, 10125 Torino (Italy) \\
$^2$Dipartimento di Fisica, Universit\`a di Genova, INFN and CNISM,
via Dodecaneso 33, 16146 Genova (Italy) \\
$^{(3)}$ CNRS, Lab. J.A. Dieudonn\'e UMR 6621,
Parc Valrose, 06108 Nice (France)}

\date{\today}

\begin{abstract}
We study the effects of polymer additives 
on turbulence generated by the ubiquitous Rayleigh-Taylor 
instability.   
Numerical simulations of complete viscoelastic models provide 
clear evidence that the heat transport is enhanced up to $50 \%$ 
with respect to the Newtonian case. This phenomenon is accompanied
by a speed up of the mixing layer growth. 
We give a phenomenological interpretation of these results based on 
small-scale turbulent reduction induced by polymers.
\end{abstract}


\maketitle

Controlling transport properties in a turbulent flow is an issue of 
paramount importance in a variety of situations ranging from pure sciences 
to technological applications \cite{G2000,siggia_arfm94,WMD01}.
After Toms \cite{toms49}, one of the most spectacular way to achieve
this goal consists in adding inside the fluid solvent a small amount 
of long-chain polymers (parts per million by weight).
The resulting fluid solution acquires a non-Newtonian character and the 
most interesting dynamical effect played by polymers is encoded in the 
drag coefficient,
a dimensionless measure of the power needed to maintain a
given throughput in a pipe. With respect the Newtonian case 
(i.e., in the absence of polymers), it 
can be reduced up to $80\%$ \cite{lumley69,sw_jfm00}. 

In many relevant situations ({\it e.g.} atmospheric convection) 
the velocity field is two-way coupled to the temperature field 
with the result that, together with mass, also heat is transported 
by the flow. 
Because drag reduction is associated to mass transport enhancement,
an intriguing question is on whether this is accompanied by a similar
variation in the heat transport.

\begin{figure}[htb!]
\includegraphics[clip=true,keepaspectratio,width=4.0cm]{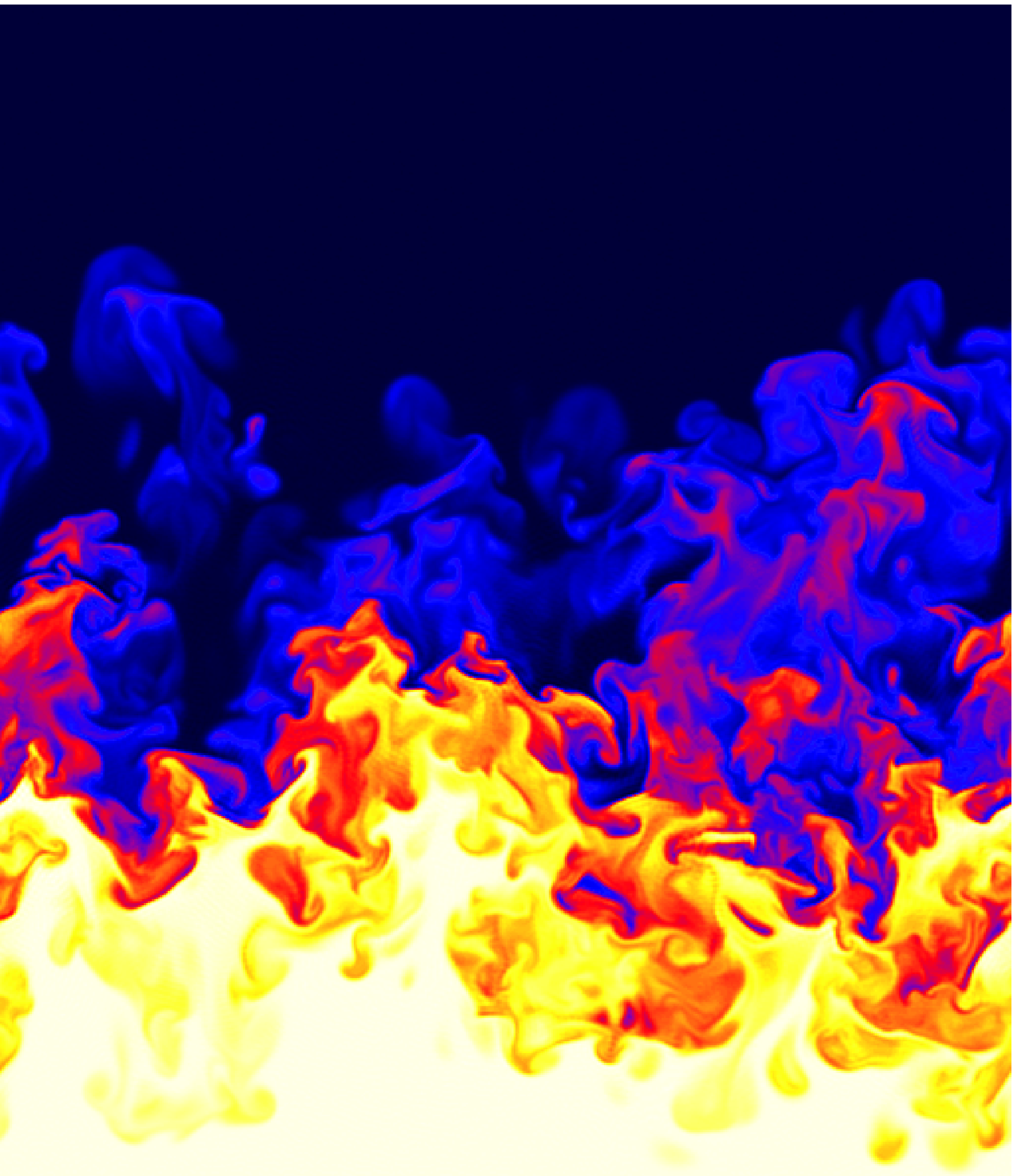}
\includegraphics[clip=true,keepaspectratio,width=4.0cm]{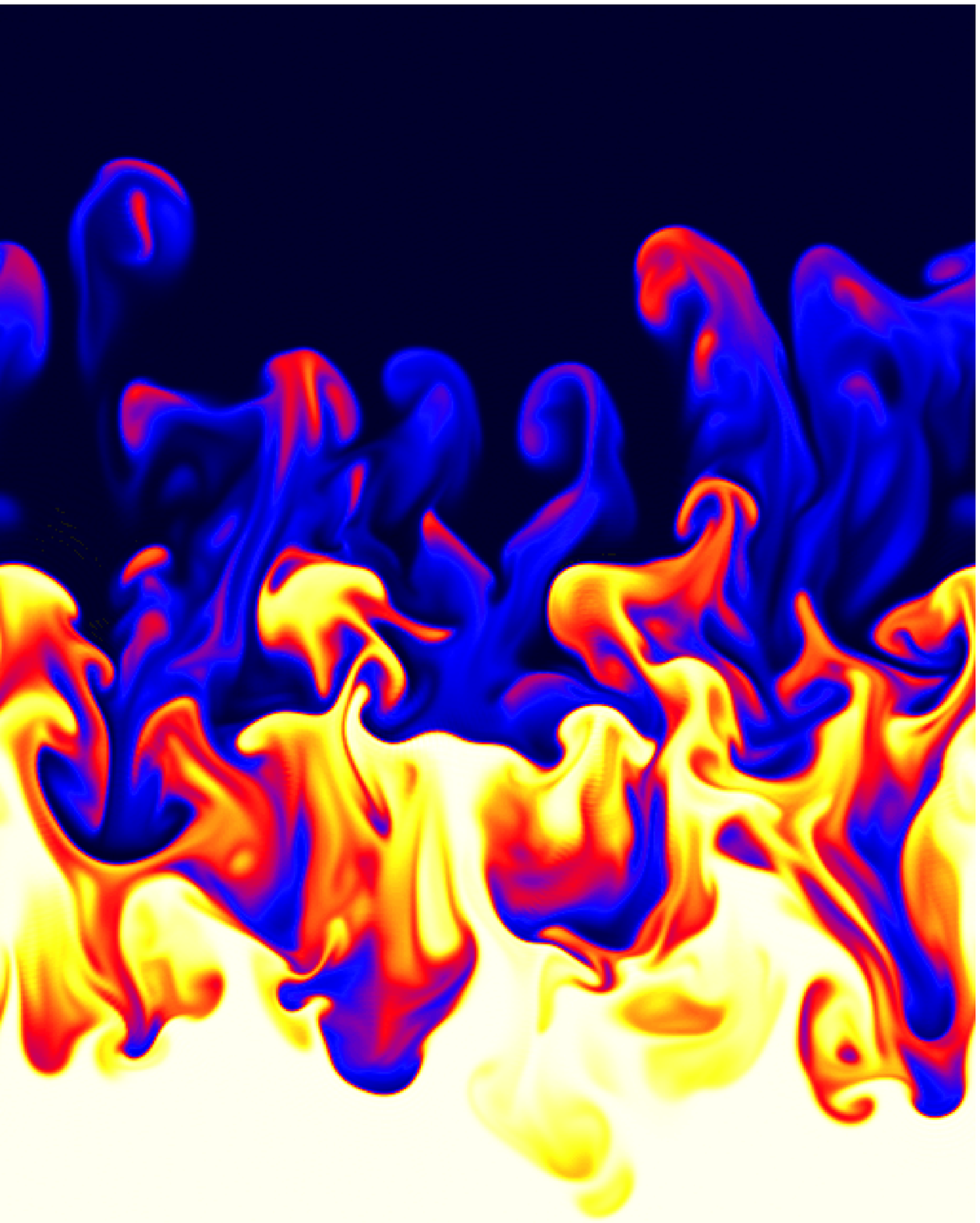}
\caption{Vertical sections of temperature field for Newtonian (left) 
and viscoelastic (right) RT simulation at time $t=2 \tau$ 
starting from the same initial conditions.
White (black) regions correspond to hot (cold) fluid.
Boussinesq-Oldroyd-B equations (\ref{eq:1}) 
are integrated by a standard fully dealiased pseudo-spectral 
code on uniform grid at resolution 
$512 \times 512 \times 1024$.
Physical parameters are $Pr=\nu/\kappa=1$, $\eta=0.2$ ($\eta=0$ for 
Newtonian run), $\beta g=0.5$, $\theta_0=1$ ($A g=0.25$).
Deborah number $De=\tau_p/\tau$ is $De=0.2$.
The initial perturbation is seeded in both cases by adding a
$10\%$ of white noise (same realization for both runs) 
to the initial temperature profile in a small
layer around the middle plane $z=0$.
}
\label{fig1}
\end{figure}

In this Letter we demonstrate the simultaneous occurrence of 
mass transport enhancement (drag reduction) and 
{\it heat transport enhancement} induced by polymers 
in a three-dimensional buoyancy driven turbulent flow originated by 
the ubiquitous Rayleigh--Taylor (RT) instability 
\cite{rayleigh_plms83,taylor_prsl50}.
This instability arises at the interface between a layer of light
fluid and a layer of heavy fluid placed above and develops
in a turbulent mixing layer (see Fig.~\ref{fig1}) 
which grows accelerated in time.
Heuristically, the RT system can be assimilated to a channel 
inside which vertical motion of thermal plumes is maintained by the 
available potential energy.
Our idea on the possibility of observing drag reduction in this 
system is suggested by recent analytical results 
which show a speed-up of the instability due to polymer additives
\cite{bmmv_jfm10}. Moreover, examples of turbulent drag reduction 
without boundaries have been recently 
provided, {\it e.g.}, in \cite{bdgp_pre03,dcbp_jfm05,bbbcm_epl06,bcm05} 

Direct numerical simulations of primitive
equations show that thermal plumes are faster in the presence
of polymers (see Fig.~\ref{fig1}), therefore the mixing layer 
accelerates (up to $30\%$ at final observation time) with respect to the Newtonian 
case and complete mixing is achieved in a shorter time.
A second and more dramatic effect, also clearly detectable in Fig.~\ref{fig1},
is that polymers reduce small scale turbulence 
\cite{bdgp_pre03,dcbp_jfm05,bbbcm_epl06}. As a 
consequence, thermal plumes in the viscoelastic case are more coherent
and transport heat more efficiently. Quantitatively, the enhancement
of the heat transport corresponds to larger values (more than $50 \%$ at
final observation time) of the Nusselt number with respect the Newtonian case.

We consider the incompressible RT system  in the 
Boussinesq approximation generalized to a viscoelastic fluid using
the standard Oldroyd-B model \cite{bhac87}
\begin{equation}
\begin{array}{lll}
\partial_t {\bf u} + {\bf u} \cdot {\bf \nabla} {\bf u} & = & 
- {\bf \nabla} p + \nu \nabla^2 {\bf u} - \beta {\bf g} T + 
{2 \nu \eta \over \tau_p}
{\bf \nabla} \cdot \sigma  \\
\partial T + {\bf u} \cdot {\bf \nabla} T & = &  \kappa \nabla^2 T \\
\partial_t \sigma + {\bf u} \cdot {\bf \nabla} \sigma & = & 
({\bf \nabla} {\bf u})^{T} \cdot \sigma + \sigma \cdot 
({\bf \nabla} {\bf u}) - {2 \over \tau_p} (\sigma - \mathbb{I}) 
\end{array}
\label{eq:1} 
\end{equation}
together with the incompressibility condition 
${\bf \nabla} \cdot {\bf u} = 0$.
In (\ref{eq:1}) $T({\bf x},t)$ is the temperature field, proportional to
the density via the thermal expansion coefficient $\beta$ as
$\rho=\rho_0 [1-\beta (T-T_0)]$ ($\rho_0$ and $T_0$ are reference
values), $\sigma_{ij}({\bf x},t)$ is the
positive symmetric conformation tensor of polymer molecules, 
${\bf g}=(0,0,-g)$ is gravity acceleration, 
$\nu$ is the kinematic viscosity, $\kappa$ is the thermal diffusivity,
$\eta$ is the zero-shear polymer contribution to viscosity
(proportional to polymer concentration)
and $\tau_p$ is the (longest) polymer relaxation time \cite{bhac87}.

The initial condition for the RT problem is an unstable temperature
jump $T({\bf x},0)=-(\theta_0/2)sgn(z)$ in a fluid at rest
${\bf u}({\bf x},0)=0$ and coiled polymers $\sigma({\bf x},0)=\mathbb{I}$.
The physical assumptions under which the set of equations (\ref{eq:1}) is
valid are of small Atwood number $A=(1/2) \beta \theta_0$ 
(dimensionless density fluctuations) and dilute polymer solution.
Experimentally, density fluctuations can also be obtained by some
additives (e.g., salts) instead of temperature fluctuations: within the
validity of Boussinesq approximation, these situations are described 
by the same set of equations (\ref{eq:1}).
In the following, all physical quantities are made dimensionless using
the vertical side, $L_z$, of the computational domain, 
the temperature jump $\theta_0$ and the
characteristic time $\tau=(L_z/A g)^{1/2}$ as fundamental units.
Elasticity of the polymer solution is measured by the Deborah number $De$,
the ratio of polymer relaxation time to a characteristic time of the flow. 
In our unsteady case $De$ grows in time starting from $De=0$, therefore
viscoelastic effects are initially absent.
An estimate of the largest Deborah number achievable is 
based on the large scale convective time as $De=\tau_p/\tau$.

\begin{figure}[htb!]
\includegraphics[clip=true,keepaspectratio,width=8.5cm]{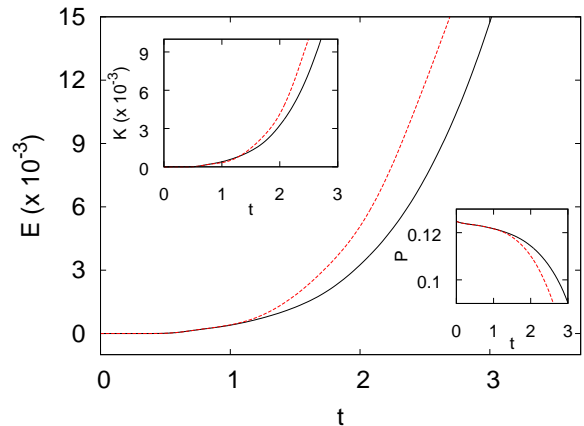}
\caption{Time evolution of total energy $E=K+\Sigma$, kinetic energy
$K$ (upper inset) and potential energy $P$ (lower inset) for the Newtonian 
run ($De=0$, black continuous line) and the viscoelastic run 
($De=0.2$, red dashed line).
}
\label{fig2}
\end{figure}

Total energy of the solution has an additional elastic contribution
to kinetic energy 
$E=K+\Sigma=(1/2) \langle u^2 \rangle + (\nu \eta/\tau_p) \langle tr 
\sigma \rangle$ and the energy balance for (\ref{eq:1}) reads
\begin{equation}
- {d P \over dt} = \beta g \langle w T \rangle = 
{d E \over dt} + \varepsilon_{\nu} + {2 \nu \eta \over \tau_p^2}
\left[ \langle tr \sigma \rangle - 3 \right]
\label{eq:2}
\end{equation}
where $P=-\beta g \langle z T \rangle$ is the potential energy and
$\varepsilon_{\nu}=\nu \langle (\partial_{\alpha} u_{\beta})^2 \rangle$
is the viscous dissipation and the last term represents elastic dissipation.
Because this last term in (\ref{eq:2}) is not negative, one might expect that 
the presence of polymers accelerates the consumption of 
potential energy with respect to the Newtonian case ($\eta=0$),
as it is indeed observed in Fig.~\ref{fig2}.

Of course, the speed-up of potential energy consumption due to polymers
does not automatically imply the increase of kinetic energy growth.
Part of potential energy is indeed converted to elastic energy $\Sigma$ by 
polymers elongation.
The inset of Fig.~\ref{fig2} shows indeed that kinetic energy for 
viscoelastic runs is larger than in the Newtonian case 
(of about $40 \%$ at $t=2.5 \tau$). 
This is the fingerprint of a ``drag reduction'' as defined for 
homogeneous-isotropic turbulence in the absence of a mean flow 
\cite{bdgp_pre03,dcbp_jfm05}.

\begin{figure}[htb!]
\includegraphics[clip=true,keepaspectratio,width=8.5cm]{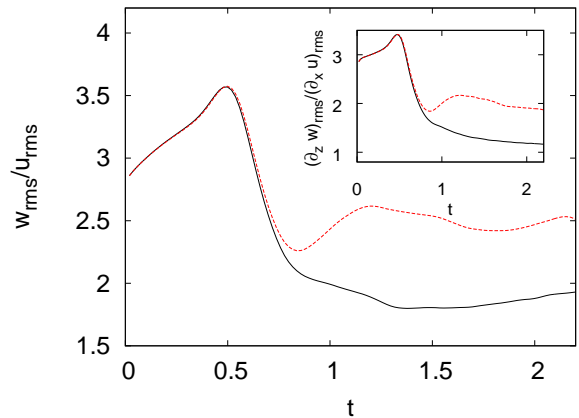}
\caption{Time evolution of velocity anisotropy $w_{rms}/u_{rms}$ 
where $w$ and $u$ are the vertical and one horizontal velocity components,
respectively. Black line is the Newtonian run at $De=0$, 
red dashed line is the viscoelastic run at $De=0.2$.
Inset: the evolution of the ratio of
velocity gradients $(\partial_z w)_{rms}/(\partial_x u)_{rms}$.
}
\label{fig3}
\end{figure}

The most important effect of polymers on turbulent velocity is 
to generate more coherent thermal plume with respect the Newtonian
case, as it is evident in Fig.~\ref{fig1}. 
This reflects in larger vertical component of the velocity 
with respect the horizontal one, {\it i.e.}, an increased anisotropy
of the velocity field. This effect is evident in Fig.~\ref{fig3} where
we plot the ratio of vertical rms velocity $w_{rms}$ to horizontal
one $u_{rms}$. The anisotropy ratio, which is around $1.8$ for the 
Newtonian case \cite{bmmv_pre09}, becomes larger than $2.5$ for the
viscoelastic run.
More important, in the viscoelastic case the
anisotropy persists also at small scales ({\it i.e.}, in the ratio of
velocity gradients), while it is almost absent in the Newtonian case
(see inset of Fig.~\ref{fig3}). 

\begin{figure}[htb!]
\includegraphics[clip=true,keepaspectratio,width=8.5cm]{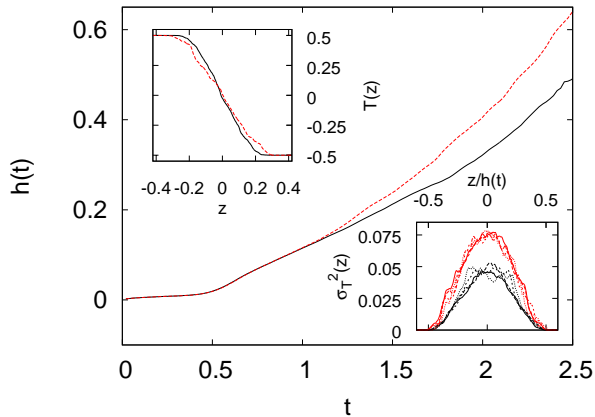}
\caption{Growth of the mixing layer thickness $h_{0.98}(t)$ defined as
the vertical range for which $|\overline{T}(z)| \le 0.98 \theta_{0}/2$
as a function of dimensionless time $t/\tau$ for Newtonian run
(black line) and the viscoelastic run 
(dashed red line) starting from the same initial condition.
Upper inset: mean temperature profiles $\overline{T}(z)$ for the two
cases at time $t=2 \tau$. Lower inset: temperature variance profiles
$\sigma_{T}^{2}(z)$ at different times
vs $z/h(t)$ (black Newtonian, red: viscoelastic run).
}
\label{fig4}
\end{figure}

Despite the fact that RT turbulence has vanishing mean flow,
a natural mean velocity is provided by the growth of the
width $h(t)$ of the turbulent mixing layer where heavy and light
fluids are well mixed.
For ordinary fluids at small viscosity, 
as a consequence of constant acceleration, 
one expects $h(t)= \alpha A g t^2$ where $\alpha$ is a dimensionless
parameter to be determined empirically
\cite{ra_pof04,dimonte_etal_pof04,krbghca_pnas07}.
Several definitions of $h(t)$ have been proposed, based on 
either local or global properties of the mean temperature profile
$\overline{T}(z,t)$ (the overbar indicates average over the horizontal
directions) \cite{as_pof90,dly_jfm99,cc_natphys06,vc_pof09}.
The simplest measure $h_r$ is based on the threshold value of $z$ at which
$\overline{T}(z,t)$ reaches a fraction $r$ of the maximum value i.e.
$\overline{T}(\pm h_r(t)/2,t)= \mp r \theta_0/2$.

Figure~\ref{fig4} shows the growth of the mixing layer thickness for both
Newtonian and viscoelastic RT turbulence. 
As already suggested by Fig.~\ref{fig1}, in the viscoelastic solution
the growth of the mixing layer is faster than in the Newtonian case
({\it i.e.} larger $h(t)$, up to $30 \%$ at $t=2.5 \tau$), 
therefore we have an effect of {\it polymer drag reduction}, i.e.
polymer addiction makes the transfer of mass more efficient.
The inset of Fig.~\ref{fig4} shows that the increased efficiency is a 
global property of the mixing layer and the temperature profile of
the viscoelastic solution corresponds to the profile of the Newtonian
case at a later time. 

Also in Fig.~\ref{fig4} we plot the variance profiles
of temperature field computed at different times for both Newtonian and
viscoelastic turbulence. In both cases, $\sigma_{T}^2(z)$ at different times
collapse when plotted as a function of rescaled variable $z/h(t)$.
Therefore as turbulence develops in the domain, the level of temperature
fluctuations within the mixing layer remains constant as a consequence
of new fluctuations introduced by plumes entering from unmixed 
regions. As Fig.~\ref{fig4} indicates, the level of fluctuations 
is larger in the viscoelastic case, as a consequence of the reduced
mixing at small scales (already observed in Fig.~\ref{fig1}).
We remark that all together these results are consistent with 
the accepted phenomenology of viscoelastic homogeneous-isotropic 
turbulence where polymers simultaneously reduce energy at small scales 
and enhance energy contain at large scales
\cite{bdgp_pre03,dcbp_jfm05,bbbcm_epl06}.

The turbulent mixing layer is responsible for the huge enhancement of 
the heat exchange with respect to the steady conductive case. 
The dimensionless measure of the heat transport efficiency is usually 
given by the Nusselt number
$Nu=\langle w T \rangle h/(\kappa \theta_0)$, the ratio between
convective and conductive heat transport.
For a convective flow in the fully developed turbulent regime, the Nusselt 
number is expected to behave as a simple scaling law with respect to  
the dimensionless temperature jump which defines 
the Rayleigh number $Ra=Agh^3/(\nu \kappa)$ \cite{gl_jfm00}. 
For a flow in which boundary layers are irrelevant,
as in our case, Kraichnan predicted many years ago the so-called
ultimate state of thermal convection for which (a part logarithmic 
corrections) \cite{kraichnan62,gl_jfm00}
\begin{equation}
\begin{array}{c}
Nu = C Pr^{1/2} Ra^{1/2} \\
Re = D Pr^{-1/2} Ra^{1/2}
\end{array}
\label{eq:3}
\end{equation}
where $C$ and $D$ are numerical coefficients.
The ultimate state regime has indeed recently been observed in numerical
simulations of RT turbulence both in two and three dimensions 
\cite{cmv_prl06,bmmv_pre09,bdm_prl10}.

\begin{figure}[htb!]
\includegraphics[clip=true,keepaspectratio,width=8.5cm]{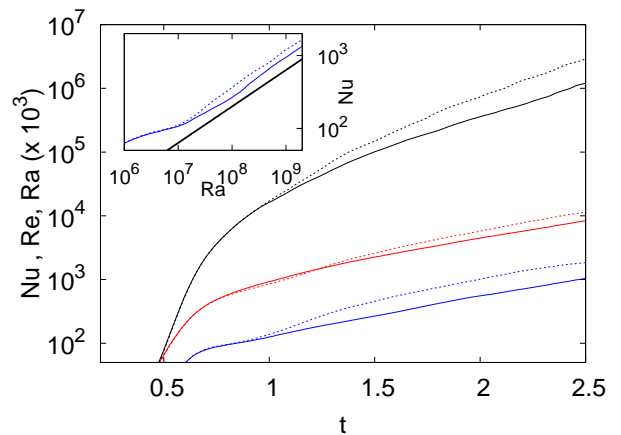}
\caption{
Time evolution of Nusselt number 
$Nu=\langle w T \rangle/(\kappa \theta_0)$ (blue, lower lines),
Reynolds number $Re=u_{rms}h/\nu$ (red, intermediate lines) and
Rayleigh number $Ra=Agh^3/(\nu \kappa)$ (black, upper lines)
for the Newtonian run
(continuous lines) and the viscoelastic run (dotted lines).
}
\label{fig5}
\end{figure}
In Fig.~\ref{fig5} we show the evolution of the Rayleigh number $Ra$, 
the Nusselt number $Nu$ and the Reynolds number $Re=u_{rms}h/\nu$ as
a function of time. For $t \ge \tau$, when turbulence is developed, all
these dimensionless quantities grow following dimensional predictions,
{\it i.e.} $Ra \sim t^6$ and $Nu \sim Re \sim t^3$. Moreover, it is 
evident that the effect of polymers is to increase the values attained
by those quantities at late time. Of course, most of this effect is due
to the enhanced value, for the viscoelastic solution, of the width 
$h(t)$ of the mixing layer which enters in the definition
of all the quantities. As discussed before, another effect
induced by polymers  is the reduction of small-scale turbulence in the
thermal plumes, which leads to an additional enhancement for the heat
flux $\langle w T \rangle$.
Therefore, the Nusselt number for viscoelastic turbulence is expected to
increase with respect to the Newtonian case when it is observed 
both as a function of  time and  as a function of  $Ra$. 
Indeed, as shown in the inset of Fig.~\ref{fig5}, both in the 
Newtonian and in the 
viscoelastic cases, $Nu \simeq Ra^{1/2}$ in agreement with the 
ultimate state regime (\ref{eq:3}) but with different coefficients, 
$C_{N}=0.022 \pm 0.002$ and $C_{VE}=0.028 \pm 0.002$ respectively,
corresponding to an increases of $27 \%$.

In conclusion, we have exploited  high resolution 
direct numerical simulations to investigate 
the effects of polymer additives on Rayleigh-Taylor
turbulence. 
There are several advantages in using the present buoyancy-driven 
turbulence system.
The presence of a time evolving mixing layer allows us 
to quantify the acceleration induced by polymers on a natural (nonzero) mean 
velocity (the mixing layer growth velocity)
at fixed buoyancy forcing, exactly
in the same spirit of usual drag reduction in bounded flows.
The relative simple and well understood phenomenology of the heat
transport (which follows the Kraichnan's ultimate state regime)
allows us to quantify the effects of polymers on the heat transport.
While the former feature is specific of the present configuration, the 
latter occurs in the bulk of the mixing region and therefore 
we conjecture that our findings hold in situations more general
than the specific setup we studied, as indeed a recent investigation seems
to indicate \cite{bcd_prl10}. 
Moreover, RT turbulence can be realized in laboratory experiments
and therefore our results based on numerical simulations of primitive
equations are a starting point of the experimental investigation of 
polymer additive effects on buoyancy-driven turbulent systems.

We thank the Cineca Supercomputing Center (Bologna, Italy) for the
allocation of computational resources.

\bibliography{biblio2}{}

\end{document}